\documentclass[preprint,12pt,3p]{elsarticle}

\usepackage{amssymb}
\usepackage{graphicx}
\usepackage[cmex10]{amsmath}
\usepackage{array}
\usepackage{mdwmath}
\usepackage{caption}

\journal{X}

\begin{document}

\begin{frontmatter}

\title{Three Shades of Partial Protection in Elastic Optical Networks}

\author[label1]{Dao Thanh Hai \corref{cor1}}
\address[label1]{Post and Telecommunications Institute of Technology, Hanoi 100000, Vietnam}

\cortext[cor1]{I am corresponding author}

\ead{haidt102@gmail.com}



\begin{abstract}

Gone have been the days when the allocation of spectrum resources for a demand is over-provisioned due to the transmission quality of the worst-case demand in a network and the optical transmission technologies are based on the fixed spectral grid, modulation format and/or line-rate schemes. Elastic optical networks technologies featuring adaptive spectrum allocation mechanisms have been becoming the architecture and technology of choice for future core networks to meet explosive traffic growths in a cost and energy-efficient manner. Partial protection strategies based on the observation that in failure events, a service can tolerate a certain amount of degradation and therefore by reducing the protection traffic in the network, better spectrum utilization could be attained. Such concept has been widely studied in the traditional WDM context and yet has been somehow faded due to the fact that fixed transmission technologies allow small room for spectral improvement. This paper aims to renew the interest in partial protection and re-adapt it to the context of elastic optical networks. The evolutionary perspective we lay out in this paper identifies a new route for achieving greater spectral efficiency in a pragmatic way by simply differentiating the protection services for each demand rather than the uniform treatment for all demands. In doing so, we present a new research problem entitled, routing, modulation level, spectrum and protection service assignment which is an extension of the well-established one, that is, routing, modulation level, and spectrum assignment as the (partial) protection service for each demand is taken into account and optimized. Three variants of that problem reflecting shades of applying partial protection are covered in details. Specifically, the first one considers the intuitive case as the relative amount of protection traffic for each demand is given as the input to the network planning process while the second one is dedicated to the special case of enforcing the same figure of partial protection for all demands. More interesting is brought in the third variant where given the service level agreement for the total network traffic, we provide the optimal solution that determine the protection service for each individual demand so as to minimize the spectral occupancy. Extensive numerical simulations have been applied to the realistic COST239 network topology and it has been revealed that a remarkable spectral saving could be achieved thanks to introducing partial protection services.

\end{abstract}

\begin{keyword}
Elastic optical network \sep dedicated protection \sep fiber-optic communications \sep elastic transponders \sep partial protection \sep routing and spectrum assignment \sep network optimization \sep integer linear programming 
\end{keyword}

\end{frontmatter}


\section{Introduction}
\label{sec1}
The era of Artificial Intelligence and Big Data Analytics have been coming into play, marking a new revolution, widely referred as the Fourth Industrial Revolution (IR 4.0) \cite{ir4}. On one hand, the physical world have been progressively evolving towards digitization thanks to the convergence of several technological advances including notably sensor technologies and ubiquitous access to the Internet. Such digital transformation therefore heralds for an always-connected world in which the digital shadow of the physical world has been practically created. On the other hand, the introduction of novel and bandwidth-hungry services such as tele-presence holography and tactile control of robotics have been accelerating in an unprecedented manner. Such two driving forces have been push-pulling the scale of challenges posed by global consumption of data and the explosive growth of Internet traffic \cite{crunch}. According to the recently released report from Cisco \cite{Cisco20}, annual global IP traffic will reach 4.8 ZB per year by 2022, exhibiting a three-fold increase over a span of 5-year and such multiplicative growth has shown no signs of stopping. In this context, optical core networks forming the backbone of Internet infrastructure have been under the critical pressure for a radical re-consideration across different phases ranging from designing, and planning to operation and management to achieve greater capital and operational efficiency. Indeed, the ultimate goal is to transfer more information at a lower cost on less spectrum resources and doing so helps to produce low-latency and high-throughput backbone networks, enabling the so-called global connectivity at scale. \\

In facing with the unprecedented traffic growth, optical transport networks have been advancing accordingly. On one hand, from the architectural perspective, core optical networks have been evolving from the opaque mode to translucent and eventually fully transparent operation. Thanks to the enormous advancements in optical components, transmission technologies and photonic switching, the vision of all-optical/transparent core networks have been experimentally and practically realized, bringing in significant savings of cost, footprint, and power by eliminating unnecessary O-E-O regenerations \cite{all-optical, efficient, 20years}. On the other hand, driven by the fact that the spectrum is limited and therefore, the capacity limit of conventional fiber might soon be reached (i.e., fiber capacity crunch), elastic optical networks technologies have been proposed and developed with the objective of using fiber capacity more efficiently. In particular, thanks to significant progress in optical transmission technologies, rate-adaptive optics has been emerging as promising solution to meet ever growing bandwidth demands and simultaneously reduce network cost. In EONs, the spectrum is divided into slices, breaking the traditional rigid frequency grid in WDM networks and hence, paving the way for adaptive spectrum provisioning tailoring to specific demand including its bandwidth requirement and transmission quality \cite{EON, EON3, EON4}. On another front, as optical transport networks have been the key enabler for information society, the issue of resilience has become of greater importance \cite{Amazon}. In practice, securing the core networks against single link failures have been an integral part of network planning process and among survivability schemes, dedicated path protection has been widely deployed thanks to its rapid recovery speed and ease of operations \cite{hai_csndsp, dedicated2}. \\

Planning EONs with survivability requirements has been the focus of extensive research works, covering a wide range of protection schemes \cite{hai_ps2, hai_wiley, 1+1rsa}. It is evident that as far as protection services are provided, a certain amount of redundant capacity must be required and to optimize such redundancy, network coding has been used recently for protection purposes \cite{eon-new1, hai_comletter, hai_systems}. The majorities of these work have nevertheless relied on the conventional concept of full protection service for all demands. However, in practice, such over-provisioning appears to miss the heterogeneous nature of services. Indeed, it is well-known that roughly $25\%$ traffic traveling in optical fiber belongs to premium class while $75\%$ remaining are best-effort type and hence can be temporarily discarded on occasion of failures \cite{sla1, sla2, sla3}. Moreover, it appears that link failures are typically fixed quickly and the customers may be willing to tolerate a decreased rate if it is accompanied by reduced cost. This means that instead of providing full protection, just some part of the traffic could be delivered to receivers upon failure events and by reducing the amount of protection requirement, less spare resources might be needed and this brings to the reduction of power consumption and cost as well \cite{sla1}. This background sets the stage for the arrival of a concept, entitled, partial protection (PP) in which the differentiation of protection level for demands are taken into account to further improve network performances. Partial protection had been raised in the context of traditional WDM networks with pioneering works from \cite{pp1, pp2}. In \cite{pp1}, the authors examined the throughput gain when applying partial protection in optical WDM Networks. In \cite{pp2}, partial protection was examined to guarantees a minimum grade of service upon a link failure by utilizing backup capacity sharing between demands and such application resulted in a remarkable reductions (e.g., $83\%$) in protection resources. The scenario of traffic grooming with sub-wavelength capacity connection was studied by leveraging partial protection and it was shown in \cite{pp9} that applying PP brought out lower blocking probability and lower network costs. Multi-path routing was suitable with PP strategies and the works in \cite{pp12} proposed an effective multi-path algorithm attempting to provision bandwidth requests and simultaneously guaranteeing the maximum partial-protection possible. A related scheme, called partial path protection were covered in \cite{pp10, pp11} where a collection of backup path were used to protect an active path. Partial protection was furthermore studied in the context of disaster-survivable networks with a pioneering work from \cite{pp3}. The use of partial protection for mixed line rates networks were investigated in \cite{pp4} and the authors took advantage of multipath routing and volume discount of higher-line-rate transponders for maximizing the network performances.  However, due to the fixed nature of WDM networks, there has been little room for further improvement with partial protection and consequently the concept of partial protection had been somehow faded. The development of elastic optical networks (EONs) with adaptive resource allocation according to specific traffic requirement and/or transmission qualities opens up therefore new opportunities for a re-consideration of partial protection. In fact, a closely related strategy is called, quality-of-service provisioning, in EONs have been studied in \cite{pp5, pp6, pp7}. In \cite{pp8}, an effective machine learning-based algorithm for identifying the most probable cause of failure impacting a given service and a re-optimization algorithm to re-route affected demands with the objective of minimizing Service Level Agreement (SLA) were proposed. Multi-path allocation has been proposed as a simple way to achieve traffic protection in different networking domains and in EONs, it was shown in \cite{pp13} that dual-path allocation assuming dynamic offered traffic outperformed traditional techniques. The works in \cite{sla3} exploited the observation that traffic traveling in fiber could be classified into the premium and best-effort type and currently premium traffic made up approximately $25\%$ total traffic, better spectrum allocation could be offered. In leveraging this idea, in our recent works \cite{qos1, qos2}, the framework for adaptive provisioning satisfying the quality-of-service of individual demands have been addressed. Our works developed an optimal network design formulation taking into account the protection level of each demand as an input to the design process and it was shown that by varying the quality-of-service provisioning, greater spectrum efficiency could be achieved. Existing works in the literature nevertheless have largely focused on the direction that the level of protection traffic is specified in advance for each demand and hence provided non-optimal solution for the network design with partial protection. Different from this perspective, we are interested in a more realistic scenario as the protection level for each demand should be considered as variables of network design process and the task is to determine the optimum protection level for each demand so as to the overall service level agreement is met while minimizing the spectrum occupancy. To the best of our knowledge, this is the first time such issue has been raised and investigated.  \\ 

This paper marks a departure from conventional full protection provisioning in elastic optical networks with a concept of partial protection scheme in order to achieve greater spectrum efficiency. Leveraging partial protection offers a new perspective to re-design EONs taking into account the diversity of protection services that a traffic demand can request instead of treating all demands with same service. To this end, we present a new research problem entitled, routing, modulation level, spectrum and protection service assignment which is an extension of the well-established one, that is, routing, modulation level, and spectrum assignment as the (partial) protection service for each demand is taken into account and optimized. Our proposal of partial protection strategies in EONs is illustrated and discussed comprehensively in Section 2. In order to optimally determine the protection service for each demand, we develop the network planning framework based on mixed integer linear programming formulation in Section 3. The important point of this formulation is the consideration of protection level for each demand as a design variable to be optimized so as to meet the overall SLA requirement and to minimize the spectrum resources. To assess the efficacy of our partial protection proposal, extensive numerical simulations have been applied on the realistic topology COST239 covering diverse scenarios and the result is reported in Section 4. Finally, the conclusion of the paper and perspectives for extension is addressed in Section \ref{sec: conclusion}.

\section{The Three Shades of Partial Protection versus Full Protection}
\label{sec: QoS}
In this part, the concept of partial protection is introduced in the context of EONs. We also highlight various shades when it comes to adapting partial protection to EONs context and compare them to the conventional full protection. In backing up our proposal for partial protection in EONs, we propose a simple cost-efficient optical node architecture for practical realization of such partial protection mechanism. \\

Let us consider a conventional network planning case with full dedicated protection as illustrated clearly in Fig. 1. In this illustrative example, supposing that there are only two traffic demands in the network, that is, the first demand is from node $A$ to node $Z$ with 100 Gbps while the second demand is from node $B$ to node $Z$ with 175 Gbps requirement. Provisioning such two demands in the context of elastic optical networks means that we must select the routing, modulation format and spectrum allocation tailored to the transmission condition of each demand.  Assuming that for the first demand, both its working path and protection path have almost the same transmission quality and therefore, make use of the same modulation format (PM) 16-QAM. For a spectrum slot of $6.25$ GHz, it hence requires two spectrum slots for both working and protection signal to support $100$ Gbps traffic (i.e., each spectrum slot with PM-16-QAM modulation is capable of carrying 50 Gbps). With respect to the second demand, supposing that due to the considerable transmission length difference between its working and protection path, its working signal and protection signal operate at different modulation formats. Specifically, its working route is shorter and hence, its working signal can be modulated on higher-order modulation format (e.g., 64-QAM) than its protection signal (e.g., 8-QAM). In order to support $175$ Gbps of the second demand, three spectrum slots are needed for the working path while the protection path have to utilize five spectrum slots. A comprehensive routing, modulation format selection and spectrum assignment for the full protection is shown in Fig. 1. The spectral cost for such full protection provisioning is $7$ spectrum slots for the entire network and $31$ units of spectrum link usage. \\

\begin{figure*}[!ht]
	\centering
	\includegraphics[width=14cm, height=8cm]{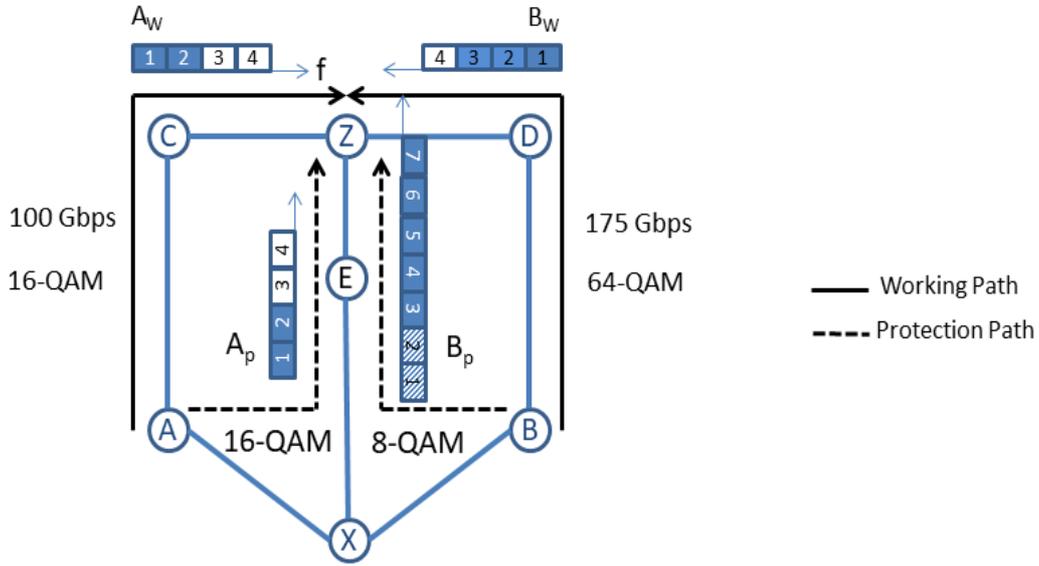}
	\caption{The traditional approach: full protection}
	\label{fig: fullprotection}
\end{figure*}

In an attempt to reduce the spectrum occupancy, we consider a different approach for offering protection service, that is, partial protection. Instead of providing full protection for all demands, let us here assume that different demands require different amount of protection traffic and that specific amount is negotiated a priori and thus becomes the inputs to the network planning process. Figure 2 illustrates the first case of partial protection where each demand is specified in advance the amount of protected traffic. Specifically, in this case, the first demand requires $50\%$ traffic upon protection mode while demand 2 needs $75\%$ amount of traffic for protection. It has been shown in Fig. 2 that doing so clearly brings about spectral savings. As far as the spectral cost is evaluated, only $5$ spectrum slots, in contrast to $7$ slots for full protection, are required for the whole network and the spectrum link usage have been reduced accordingly from $31$ units in full protection to $25$ units thanks to the reduction of protection traffic for both two demands. \\

\begin{figure*}[!ht]
	\centering
	\includegraphics[width=10cm, height=8cm]{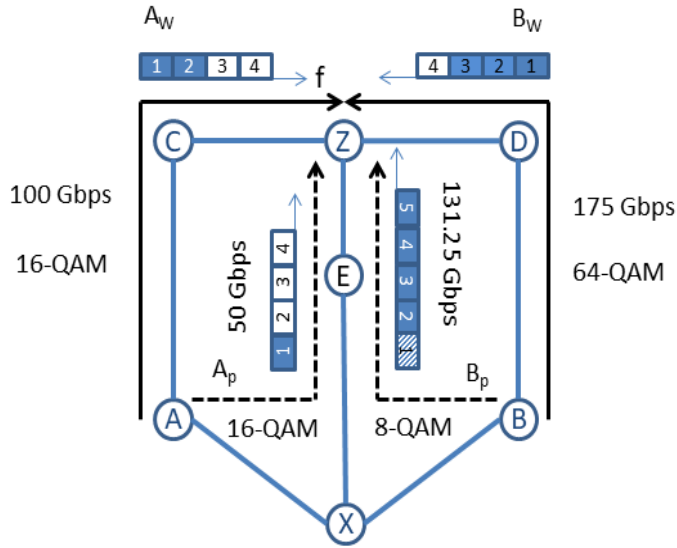}
	\caption{Partial Protection: Conventional approach}
	\label{fig: partial_c}
\end{figure*}

Let us turn the attention to a different variant of partial protection strategy where instead of specifying the protection level for each demand a priori, network operators receive the service level agreement (SLA) for the entire traffic. For example, one such SLA could be that (at least ) $75\%$ of the total traffic must be protected in case of failures. In satisfying such requirement, network operators could rely basically on two approaches. The first one is naive when all demands are treated with the same figure of protection level required by SLA. For example, given the SLA of $75\%$, all demands are offered uniformly $75\%$ amount of protection traffic. Figure 3 reveals comprehensively the application of such naive approach and it can be seen that $one$ spectrum slot is saved for the entire network compared to the full protection case. It is important to note that while shrinking from 175 Gbps to 131.25 Gbps results in $one$ spectrum slice saving for the second demand, a reduction from 100 Gbps to 75 Gbps is not significant enough to produce spectral saving for the first demand. Indeed, for demand 1, providing full protection and $75\%$ protection does not make any difference from spectral occupancy perspectives. It is observed that although the one-size-fits-all approach is intuitive, its spectral benefit is limited and hence it leads to a question of whether it exists a more spectrally efficient way to meet SLA requirement. \\

\begin{figure*}[!ht]
	\centering
	\includegraphics[width=10cm, height=8cm]{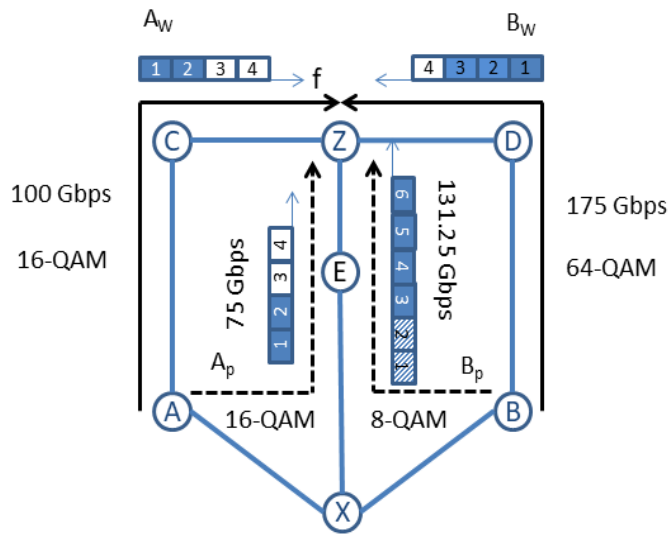}
	\caption{Partial Protection: One-size-fits-all approach}
	\label{fig: partial_a}
\end{figure*}

Considering the case that SLA of $75\%$, let us propose an alternative approach shown in Fig. 4. Instead of treating uniformly all demands with the same amount of protection level, we now vary the protection level specifically for each demand so as to the overall SLA is met. One such provisioning is illustrated in Fig. 4 as protection traffic of demand 1 is reduced by half from 100 Gbps to 50 Gbps and protection traffic of demand 2 is shrinked from 175 Gbps to 106.25 Gbps to meet SLA of $75\%$ for the entire network traffic. Overall, an amount of $75\%$ of entire traffic is secured as protection traffic. For this demand-wise approach, it is easy to verify from Fig. 4 that, $3$ spectrum slots saving could be achieved compared to the full protection scenario in Fig. 1. Moreover, it is highlighted that the demand-wise approach could generate greater spectral efficiency than the one-size-fits-all one. \\

\begin{figure*}[!ht]
	\centering
	\includegraphics[width=10cm, height=8cm]{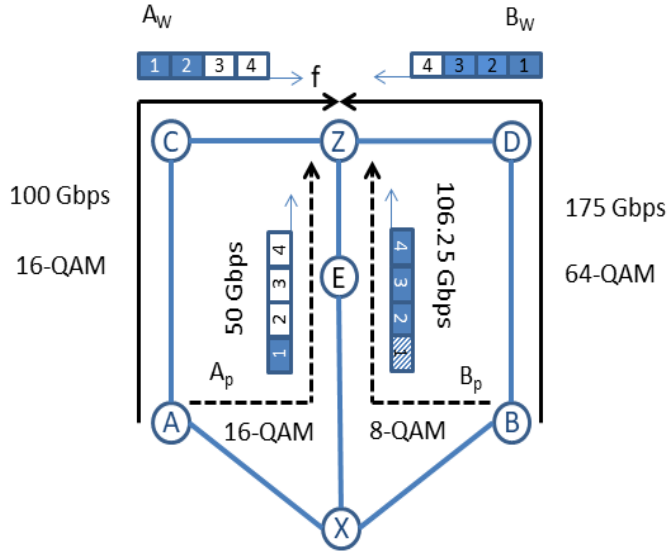}
	\caption{Partial Protection: Demand-wise approach}
	\label{fig: partial_b}
\end{figure*}

In supporting our proposed partial protection scheme, the optical node architecture has to be taken into account. Here we propose a cost-effective architecture leveraging the usage of a shared elastic transponder for both working and protection mode. The elastic transponder is capable of providing multiple modulation formats and/or baud-rate in order to support a wide range of traffic demands. The transponder is directed to the working path in the normal condition and upon the detection of failure event, the optical-layer switch is reconfigured to re-direct the transponder to the protection path. It is important to note that during this transition, the transponder might be reconfigured itself to operate at a different baud-rate and/or different modulation format to support partial protection operation. Such transition from working mode to protection mode with this node architecture has been experimentally demonstrated and it has been revealed that the overall reconfiguration could be rapidly taken place, being kept within a few tens of milliseconds satisfying the requirements of many operators \cite{review7, review9, transponder}. \\

\begin{figure*}[!ht]
	\centering
	\includegraphics[width=10cm, height=8cm]{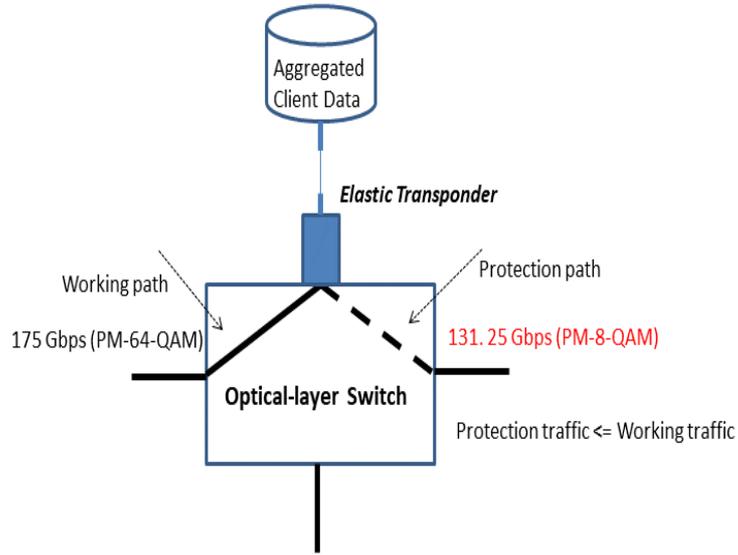}
	\caption{The optical node architecture supporting partial protection schemes}
	\label{fig: node}
\end{figure*}

The illustration provided in this part demonstrates a new venue for improving spectrum efficiency of optical networks by relying on the idea of reducing protection traffic with partial protection strategies. In general, network operators are interested in the question that given the SLA requirement for the entire network traffic, what would be the optimal assignment of protection level for each demand so that the spectral occupancy is minimized while satisfying SLA. We aim to tackle this issue in this paper and the next part provides a framework for optimal network planning with partial protection \\

\section{Mathematical Formulation for Routing, Modulation Level, Spectrum and Protection Service Assignment Problem}
\label{sec: math}
In order to realize the spectral benefits of introducing partial protection in elastic optical networks, we present a mathematical formulation for optimal network dimensioning. The formulation takes the fiber network topology, set of traffic demands and service level agreement for protection traffic as the key inputs and returns the optimum routing, modulation level, spectrum assignment for each demand together with the partial protection coefficient so as to minimize the spectrum occupancy while satisfying SLA requirement. In our formulation, the spectrum assignment with continuity and contiguity constraints are processed with the concept of pre-calculated channels \cite{modelRSA} as it was proven highly computational efficient. \\  

\noindent{Raw Input Data:}

\begin{itemize}
	\item  $G(V,E)$: The given physical topology with $|V|$ nodes and $|E|$ fiber links
	\item $S$: Set of available spectrum slices in each fiber link, indexed by $s$
	\item  $D$: Set of all traffic demands and each demand $d \in D$ requires $t^d$ Gbps
	\item $M$: Set of supportable modulation formats, indexed by $m$
	\item A function relates optical reach to modulation format and bit-rate operation $L = 145.741 + (n-14.0344) \times (60.2196 \times ln(BR) - 465.82)$: L $[km]$, n  $[bit/s/Hz]$ and bit rate (BR) $[Gbps]$ \cite{optical_reach}
	\item $K$: Set of allowable protection services, indexed by $k$. For example, $k=1$ may refer to the amount of protection of $25\%$ while $k=2$ represents to an agreement of $50\%$ protection traffic	
	\item $SLA$: Service Level Agreement specifies the overall minimum protection traffic for entire network 
\end{itemize}

\noindent{Additional Input Data obtained from Pre-processing Process:}

\begin{itemize}	 
	\item $P^{d}$: Set of $k$ pairs of link-disjoint routes for a demand $d$, each route pair $p^d$ of a demand $d$ consists of a working route and a protection route for that demand
	
	\item $P=\cup_{{d \in D}} {P^{d}}$: Set of all pairs of link-disjoint routes for all demands in the network
		
	\item $\beta_{p, k}^{d} $: a coefficient equals to 1 if a pair of routes $p \in P$ is feasible for demand $d \in D$ at protection index $k$, 0 otherwise. The pair of routes $p \in P^{d}$ for demand $d$ is considered to be feasible at protection service $k$ if there is at least one feasible modulation format carrying $t^d$ $Gbps$ traffic on the working path and there is at least one feasible modulation format carrying a partial amount of $t_d$ $Gbps$ corresponding to the protection service $k$ on protection route. 
	
	\item $h_{p, w}^{d}$: the most spectrum-efficient (highest-order) modulation format for working path belonging to a route pair $p \in P^{d}$ of demand $d \in D$
	
	\item $h_{p, b}^{d, k}$: the most spectrum-efficient (highest-order) modulation format for backup path belonging to a route pair $p \in P^{d}$ of demand $d \in D$ with respect to protection service $k$
	
	\item $C_{p, w}^d$: Set of admissible spectrum channels for working path of a route pair $p \in P^{d}$
	
	\item $C_{p, b}^{d, k}$: Set of admissible spectrum channels for backup path of of a route pair $p \in P^{d}$ with respect to protection service $k$
	
	\item $C=\cup_{{p \in P^{d}, d \in D}} {(C_{p, w}^d \cup C_{p, b}^{d, k})}$: Set of all possible spectrum channels
	
	\item $\alpha_{c, s} $: a coefficient equals to 1 if the slice $s \in S$ belongs to channel $c \in C$, 0 otherwise
	
	\item $\gamma_{p, e} $: a coefficient equals to 1 if link $e \in E$ belongs to the working path of a route pair $p \in P$, 0 otherwise
	
	\item $\delta_{p, e} $: a coefficient equals to 1 if link $e \in E$ belongs to the backup path of a route pair $p \in P$, 0 otherwise
	
	
\end{itemize}
Variables:
\begin{itemize}
	\item $x_{p, c}^{d} \in \{0,1\}$: equals to 1 if demand $d$ is provisioned by finding a pair of routes $p \in P^{d}$ and a spectrum channel $c \in C_{p, w}^d$ for working path 
	
	\item $y_{p, c} ^{d, k} \in \{0,1\}$: equals to 1 if demand $d$ is provisioned by finding a pair of routes $p \in P^{d}$ and a spectrum channel $c \in C_{p, b}^{d, k}$ for protection path with respect to protection service $k$
	 
	\item $z_k^{d} \in \{0,1\}$: equals to 1 if demand $d$ if demand $d$ is provided the protection service $k$, 0 otherwise
	 
	\item $r_{e, s} \in \{0,1\}$: equals to 1 if the spectrum slice $s \in S$ is occupied in link $e \in E$ \\
	
	\item $\theta_{s} \in \{0,1\}$: equals to 1 if the spectrum slice $s \in S$ is utilized in the network, 0 otherwise \\
	
	
\end{itemize}	

\noindent{Objective:} 
\begin{equation}
\label{eq:obj}
\textit{Minimize} \; \phi= \sum_{s \in S} {\theta_s}
\end{equation}

\noindent{Subject to the following constraints:}


\begin{equation}
\label{eq:protectionselection}
\sum_{k \in K} z^{d}_k = 1 \qquad \forall d \in D 
\end{equation}

\begin{equation}
\label{eq:partial}
\sum_{p \in {P^d}: \beta_{p}^{d, k} = 1} \sum_{c \in C_{p, b}^{d, k}} y_{p, c}^{d, k} = z^{d}_k \qquad \forall k \in K 
\end{equation}

\begin{equation}
\label{eq:dcdf}
\sum_{c \in C_{p, w}^d}x_{p, c}^d= \sum_{k \in K}\sum_{c \in C_{p, b}^{d, k}}y_{p, c}^{d, k} \qquad \forall d \in D, p \in P^d: \beta_{p}^{d, k} = 1
\end{equation}

\begin{equation}
\label{eq:sla}
\sum_{d \in D} \sum_{k \in K} z^{d}_k \times t^d \geq SLA \times \sum_{d \in D} t^d \qquad \forall d \in D 
\end{equation}

\begin{equation}
\label{eq:sliceunique}
\sum_{d \in D} \sum_{k \in K} \sum_{p \in {P^d}} \beta_{p}^{d, k} (\sum_{c \in C_{p, w}^d} x_{p, c}^d \times  \gamma_{p, e} \times \alpha_{c, s} + \sum_{c \in C_{p, b}^{d, k}} y_{p, c}^{d, k} \times \delta_{p, e} \times  \alpha_{c, s}) = r_{e, s} \qquad \forall e \in E, s \in S
\end{equation}

\begin{equation}
\label{eq:useslice}
\sum_{e \in E}{r_{e, s}}  \leq |E| \theta_s \qquad\forall s \in S 
\end{equation}

The objective in Eq. \ref{eq:obj} aims at minimizing the spectrum usage measured in terms of number of spectrum slices to support all traffic demands so as to meet the protection service level agreement for the entire traffic. Constraints in Eq. \ref{eq:protectionselection} guarantees that each demand must select one and only one protection service. The constraints in Eq. \ref{eq:partial} say that for the selected protection service of a demand $d$, a protection route and a protection channel must be found to support that demand. The coherence between the working route and protection of a demand is expressed by the constraints in Eq. \ref{eq:dcdf}. The constraints in Eq. \ref{eq:sla} are to satisfy the service level agreement for protection. The traditional constraints involving the unique usage for a spectrum slice is ensured by Eq. \ref{eq:sliceunique}. Finally, the definition of using a spectrum slice in the network, that is, a spectrum slice is considered to be used if it is occupied in any link of the network, is provided by Eq. \ref{eq:useslice}. \\   

\section{Numerical Results}	
\label{sec: results}
\begin{figure*}[!ht]
	\begin{minipage}[t]{.45\linewidth}
		\vspace{0pt}
		\centering
		\includegraphics[width=\linewidth, height=6cm]{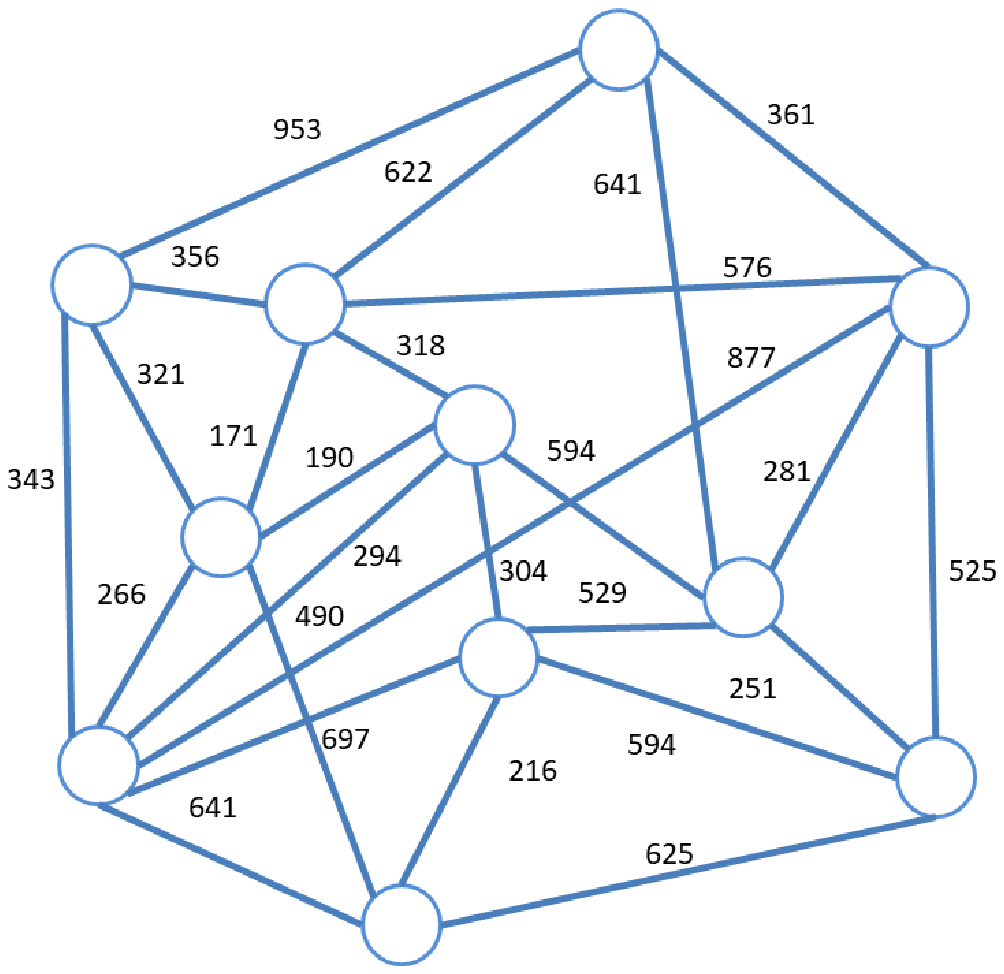}
	\end{minipage}%
	\begin{minipage}[t]{.6\linewidth}
		\vspace{2cm}
		\centering
		\begin{tabular}{|c|c|}
			\hline
			\multicolumn{2}{|c|}{\small{Descriptive Features}} \\
			\hline
			Nodes & 11  \\
			Links & 52  \\
			Average nodal degree & 4.7 \\
			Shortest link distance & $\approx 200 km$ \\
			Longest link distance & $\approx 1000 km$\\ 
			\hline
		\end{tabular}
	\end{minipage}
	\caption{COST239 network with link distance in $km$}
	\label{fig:cost239}
\end{figure*}

This section is dedicated to evaluate our proposal of partial protection on the realistic network topology COST239 as shown in Fig. \ref{fig:cost239}. This network is composed of 11 nodes and 52 fiber links whose length are ranged from $\approx$ 200 km to $\approx$ 1000 km. Such diversity in link length is a good match for the differentiation of transmission properties between the working and protection path where working signal and protection signal could be modulated at different modulation format rather than having the same one based on the worse case. For traffic generation, we assume that $50\%$ of node-pairs having the traffic and the traffic between each node pair is randomly generated between 100 Gbps and 200 Gbps with a step of 25 Gbps. Ten traffic instances are generated and six modulation formats are considered as shown in Table 1. The number of pairs of route per demand $k$ is 4 and we make use of a variant of Suurballe algorithm to find such 4 shortest pairs for a demand. The mixed integer linear programming formulation in Section 3 is applied and CPLEX solver with an academic version \cite{IBM} is used to obtain optimal results. \\

\begin{table*}[!ht]
	\caption{Bit-rate Operation per $6.25$ $GHz$ Spectrum Slot for different Modulation Formats}
	\label{tab:result0}
	\centering
	\begin{tabular}{cc}
		\hline
		Modulation Formats & Bit-rate per Spectrum Slot \\
		\hline
		PM-BPSK & 12.5 Gbps \\
		PM-QPSK & 25 Gbps \\
		PM-8QAM & 37.5 Gbps \\
		PM-16QAM & 50 Gbps \\
		PM-32QAM & 62.5 Gbps \\
		PM-64QAM & 75 Gbps \\
		\hline
	\end{tabular}
\end{table*}

For benchmarking, in the first scenario, we compare the traditional full protection design with the partial protection design where the relative amount of protection traffic for each demand is randomly selected among $25\%$, $50\%$ and $75\%$ and provided as input to the design process. Figure 7 shows the comparison of such two designs in terms of used spectrum slices. As expected, by reducing the amount of protection traffic, less spectrum slots are required to accommodate the traffic. In this simulation scenario, the result reveals that instead of over-provisioning with full protection traffic for all demands, remarkable spectral saving could be achieved, up to $32\%$ for partial protection approach \\

\begin{figure*}[!ht]
	\centering
	\includegraphics[width=14cm, height=6.5cm]{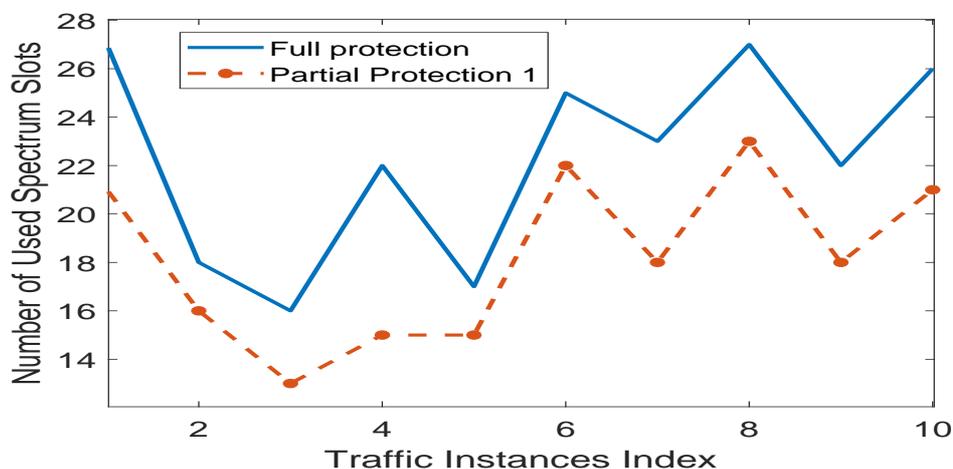}
	\caption{Full protection versus Partial Protection in terms of used spectrum slices}
	\label{fig:result1}
\end{figure*}

Let us move to the second scenario of applying partial protection. For this case, all demands are enforced to have the same protection services specified by the SLA for the entire network. Three kind of SLA protection services are considered, that is, $25\%$, $50\%$ and $75\%$. Figure 8 reports the impact of spectral occupancy as the amount of protection traffic is varied from $75\%$ to $25\%$ in comparison with the full protection. It is clearly shown that shrinking the protection trafffic generates spectral savings and specifically such saving could be up to $26\%$ for $75\%$ partial protection, $46\%$ for $50\%$ partial protection and $55\%$ for $25\%$ partial protection. It can also be observed that the spectral saving does not move proportinally to the reduction of protection traffic. In our studied case, shrinking protection traffic up to $50\%$ tends to achieve the best revenue while going beyond $50\%$ brings about relatively small impact. \\

\begin{figure*}[!ht]
	\centering
	\includegraphics[width=14cm, height=6.5cm]{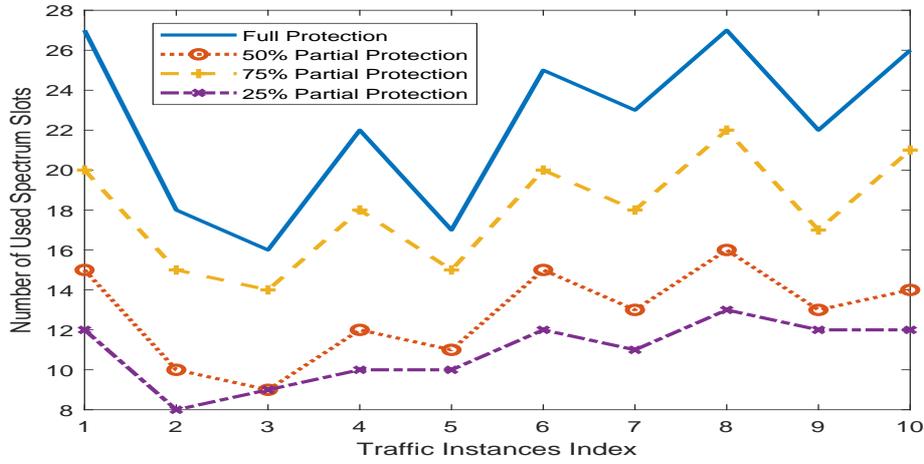}
	\caption{Full protection versus Partial Protection $25\%$, $50\%$ and $75\%$}
	\label{fig:result2}
\end{figure*}

We turn the attention to third scenario where the relative amount of protection traffic for each demand becomes a design variable and it has to be optimized to meet the SLA of entire network. We reconsider the second scenario but now instead of enforcing the same amount of protection traffic for all demands, each demand can select the protection services among three types $25\%$, $50\%$ and $75\%$ according to the optimization model in Section 3. Figure 9 compares the one-size-fits-all approach and the demand-wise one when the SLA is $50\%$. It can be clearly seen that by optimizing the selection of protection service for each demand, better spectrum usage can be realized. In our simulation of ten traffic instances, 8 out 10 cases results in greater spectrum efficiency with the demand-wise approach compared to the one-size-fits-all one and in the most favorable case, the demand-wise approach could be $25\%$ more efficient than its counterpart.  

\begin{figure*}[!ht]
	\centering
	\includegraphics[width=14cm, height=6.5cm]{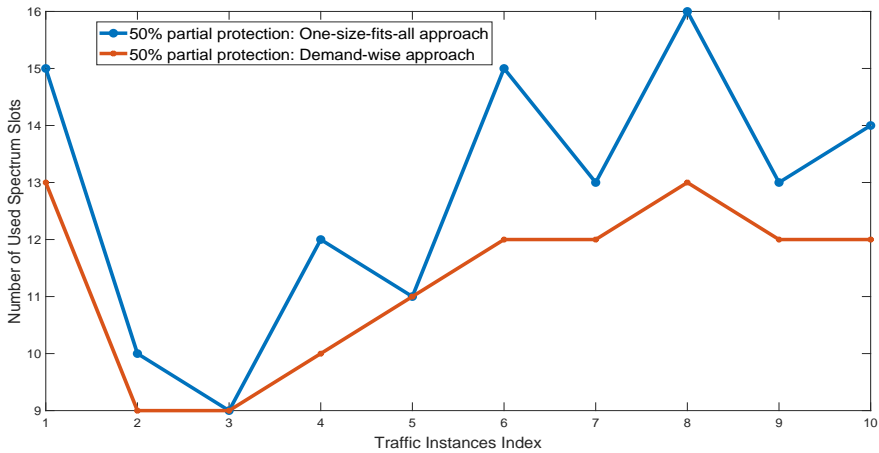}
	\caption{One-size-fits-all partial protection versus Demand-wise partial protection}
	\label{fig:result3}
\end{figure*}

.  
\section{Conclusions}
\label{sec: conclusion}
In an attempt to achieve greater capacity efficiency for optical core networks, this paper proposed the concept of partial protection to be applied in elastic optical networks, that is, instead of providing full protection for all traffic demands, the assignment of (partial) protection traffic for each demand is taken into account and optimized. We examined various scenarios that applying partial protection could potentially generate spectral savings and introduced the network design framework incorporating partial protection. A new problem, entitled, routing, modulation level, spectrum and protection service assignment was developed for optimal network designs based on mixed integer linear programming formulation. In verifying the efficacy of our proposal, extensive simulation on the realistic network topology COST239 was carried out and our findings indicated that significant savings of spectrum resources could be achieved by reducing the protection traffic. Moreover, our results demonstrated that the demand-wise approach far outperformed the one-size-fits-all and the difference could be up to $25\%$ in our studied cases. \\

\section*{Conflict of interest}
The authors declare that they have no conflict of interest.

\bibliographystyle{elsarticle-num}

\bibliography{ref}

\end{document}